\documentclass[journal]{IEEEtran}
\ifCLASSINFOpdf
  \usepackage[pdftex]{graphicx}

\usepackage{cite}
\usepackage[utf8]{inputenc}
\usepackage{csquotes}
\usepackage{amsmath}
\usepackage{xcolor}

\begin{document}

\title{High-Resolution Atomic Magnetometer-Based Imaging of Integrated Circuits and Batteries}

\author{Dominic~Hunter,
        Marcin~S.~Mrozowski,
        Stuart~J.~Ingleby,       
        Timothy~S.~Read,
        Allan~P.~McWilliam,
        James~P.~McGilligan,
        Ralf~Bauer,
        Peter~D.~D.~Schwindt,
        Paul~F.~Griffin,
        Erling~Riis%

\thanks{D. Hunter, M. S. Mrozowski, S. J. Ingleby, A. P. McWilliam, J. P. McGilligan, P. F. Griffin, and E. Riis are with the Department of Physics, SUPA, University of Strathclyde, Glasgow G4 0NG, UK (email: d.hunter@strath.ac.uk; marcin.mrozowski@strath.ac.uk; stuart.ingleby@strath.ac.uk; allan.mcwilliam@strath.ac.uk; james.mcgilligan@strath.ac.uk; paul.griffin@strath.ac.uk; e.riis@strath.ac.uk).}
\thanks{R. Bauer is with the Department of Electronic \& Electrical Engineering, University of Strathclyde, Glasgow G1 1XW, UK (email: ralf.bauer@strath.ac.uk).}%
\thanks{P. D. D. Schwindt is with Sandia National Laboratories, Albuquerque, NM 87185, USA (email: pschwin@sandia.gov).}
\thanks{T. S. Read is with the Center for Quantum Information and Control, Department of Physics \& Astronomy, University of New Mexico, Albuquerque, NM 87106, USA, and also with Sandia National Laboratories, Albuquerque, NM 87185, USA (email: tsread@sandia.gov).}
}

\maketitle


\begin{abstract}
Optically pumped magnetometers (OPMs) have emerged as a powerful technique for high-resolution magnetic field imaging. However, achieving sub-millimeter spatial resolution at sub-picotesla sensitivities ($\mathbf{<1}$~pT/$\boldsymbol{\sqrt{\mathrm{Hz}}}$) remains challenging, particularly under finite-field conditions. We present a high-resolution magnetic imaging system based on a free-induction-decay (FID) OPM integrated with a two-axis scanning micromirror for automated beam steering. The double-pass optical configuration allows millimeter-scale devices under test (DUTs) to be positioned directly behind the vapor cell. This enables a standoff distance of 2.7~mm between the magnetic source and the atomic vapor, improving practical imaging resolution by increasing the amplitude of near-field magnetic signals sampled within the sensitive volume. Spatial resolution is experimentally demonstrated by imaging a custom printed circuit board (PCB) containing antiparallel copper tracks spaced 2~mm apart, with measured field maps in close agreement with Biot--Savart predictions. The OPM achieves an optimal field sensitivity of $\mathbf{0.5}$~pT/$\boldsymbol{\sqrt{\mathrm{Hz}}}$, demonstrating the system’s capability for high-precision magnetic field measurements. The imaging system is further validated by resolving polarity-dependent asymmetries in a bridge rectifier integrated circuit (IC) and tracking current dynamics in a ceramic battery \textit{in situ}. These results highlight the potential of OPM-based systems for noninvasive diagnostics of electronic circuits and batteries.
\end{abstract}


\section{Introduction}
\IEEEPARstart{M}{agnetic} field imaging is a powerful diagnostic tool with applications in biomedical science~\cite{zhao2023optically}, integrated circuit (IC) analysis~\cite{kehayias2022measurement}, industrial monitoring~\cite{bai2023atomic}, and security~\cite{krelina2021quantum}. These diverse use cases share a common requirement: the ability to resolve spatial and temporal variations in magnetic signatures with high precision. Among available sensing technologies, optically pumped magnetometers (OPMs) have emerged as leading candidates owing to their exceptional sensitivity while operating under ambient conditions and maintaining low size, weight, and power (SWaP) characteristics~\cite{ingleby2022digital, limes2020portable}. \\
\indent Despite the excellent field sensitivity achievable with optically pumped magnetometers (OPMs), the spatial resolution of a single sensor module is influenced by several factors, including the size of the optical interrogation volume, atomic spin diffusion within the vapor cell, and the standoff distance between the sensing volume and the magnetic source \cite{horsley2015high}. In practical implementations, spatial resolution is often primarily limited by the standoff distance, which is constrained by cell packaging, device dimensions, thermal insulation, and optical components that restrict proximity to the device under test (DUT). As a result, OPMs typically achieve standoff distances no better than several millimeters when operated as standalone sensors~\cite{nordenstrom2024feasibility}. \\
\indent Several approaches have been explored to mitigate these limitations. For example, nitrogen-vacancy (NV) center magnetometry in diamond can attain nanoscale spatial resolution~\cite{garsi2024three}, but with limited magnetic precision~\cite{ariyaratne2018nanoscale}. Radio-frequency alkali-vapor magnetometers~\cite{bevington2019imaging} and electromagnetic induction imaging schemes~\cite{maddox2023rapid} improve spatial sampling through raster-scanned interrogation, but often suffer from limited dynamic range or require stringent magnetic-field compensation. In biomedical contexts, multi-sensor OPM arrays~\cite{alem2017magnetic, boto2018moving} enhance spatial coverage and noise rejection; however, their achievable spatial resolution remains constrained by the physical size and spacing of individual sensor modules. More recently, multi-channel configurations based on a single vapor cell~\cite{kim2014multi} have demonstrated improved spatial resolution and common-mode noise suppression, albeit at the cost of increased optical and system complexity. Consequently, achieving sub-millimeter spatial resolution while maintaining high sensitivity in finite-field environments remains a critical step in expanding the utility of magnetic imaging systems~\cite{mitchell2020colloquium}. \\
\indent A promising strategy is to integrate optical beam steering within a single vapor cell, enabling high spatial resolution while retaining the impressive sensitivity of OPMs. Recent demonstrations using digital micromirror devices (DMDs) in the spin-exchange relaxation-free (SERF) regime achieved a sensitivity of $25$~fT/$\sqrt{\mathrm{Hz}}$ within a zero-field environment, at $\mathrm{216~\mu{m}}$ resolution that was defined by the DMDs segmentation of a broad optical beam~\cite{fang2020high}. Finite-field free-induction-decay (FID)-based methods have also been adopted with reduced sensitivity ($\approx 10$~pT/$\sqrt{\mathrm{Hz}}$)~\cite{liu2021submillimeter}. \\
\indent Building on these advances, this work presents a FID-based OPM imaging system that experimentally demonstrates sub-picotesla sensitivity and robust operation in finite magnetic fields, with the ability to resolve millimeter-spaced current features under the present experimental conditions while establishing a clear pathway toward sub-millimeter spatial resolution. This is achieved by positioning the DUT directly adjacent to a compact MEMS vapor cell, thereby minimizing the standoff distance to $2.7\,\mathrm{mm}$. The resulting system provides access to a previously unexplored sensing regime, extending magnetic imaging capabilities between high-precision OPM systems and nanoscale diamond-based magnetometers~\cite{mitchell2020colloquium}. \\
\indent \indent The proposed system employs a microelectromechanical systems (MEMS) vapor cell, a double-pass optical geometry, and a two-axis electrostatic gimbal-less MEMS micromirror for automated beam steering. Spatial resolution is improved by raster scanning a tightly confined optical interrogation volume across the vapor cell while minimizing the standoff distance between the sensing volume and the DUT. Optical beam steering avoids mechanical translation of the sensor head, enabling spatially resolved imaging while maintaining high magnetometer sensitivity. In the present implementation, the MEMS micromirror introduces additional vibrational noise, leading to a slight reduction in sensitivity. This technical limitation is expected to be reducible by optimizing the micromirror driving electronics. Additionally, a Hilbert transform-based digital signal processing (DSP) method~\cite{wilson2020wide, hunter2023optical} enables rapid frequency extraction from the FID signals, offering over an order-of-magnitude enhancement in processing efficiency over nonlinear fitting algorithms while preserving accuracy and precision. \\
\indent We demonstrate the system’s performance by imaging diverse magnetic sources, including printed circuit boards (PCBs), a bridge rectifier integrated circuit (IC) under different bias current conditions, and a surface-mount ceramic battery during both charging and discharging cycles. In particular, measurements of a custom PCB containing antiparallel current-carrying tracks, spaced $2\,\mathrm{mm}$ apart, provide an experimental demonstration of the system’s spatial resolution, with the measured magnetic field distributions showing close agreement with Biot--Savart predictions. These results establish the presented FID-based OPM imaging platform as a robust, scalable, and noninvasive diagnostic tool for electronics and battery monitoring in industrial environments~\cite{borna2018magnetic, bason2022non}.

\section{Methods}
This section describes the design and operation of the magnetic imaging system. Section~\ref{subsection:experimental_setup} presents the overall experimental setup and highlights key innovations relative to previous FID OPMs~\cite{hunter2023optical, hunter2023free}, including a double-pass optical geometry and MEMS-based beam steering. Section~\ref{subsection:frequency_estimation} details the Hilbert transform-based frequency extraction method developed for rapid signal processing, while Section~\ref{subsection:mems_mirror} explains the implementation and performance of the two-axis MEMS micromirror. 
\subsection{Experimental Setup}
\label{subsection:experimental_setup}
To achieve high-resolution magnetic imaging, the system integrates a miniaturized cesium (Cs) vapor cell with optical access optimized for near-field sensing. This section outlines the design considerations and configuration of the FID-based magnetometer. \\
\begin{figure}[t]
\centering
\includegraphics[width=\columnwidth]{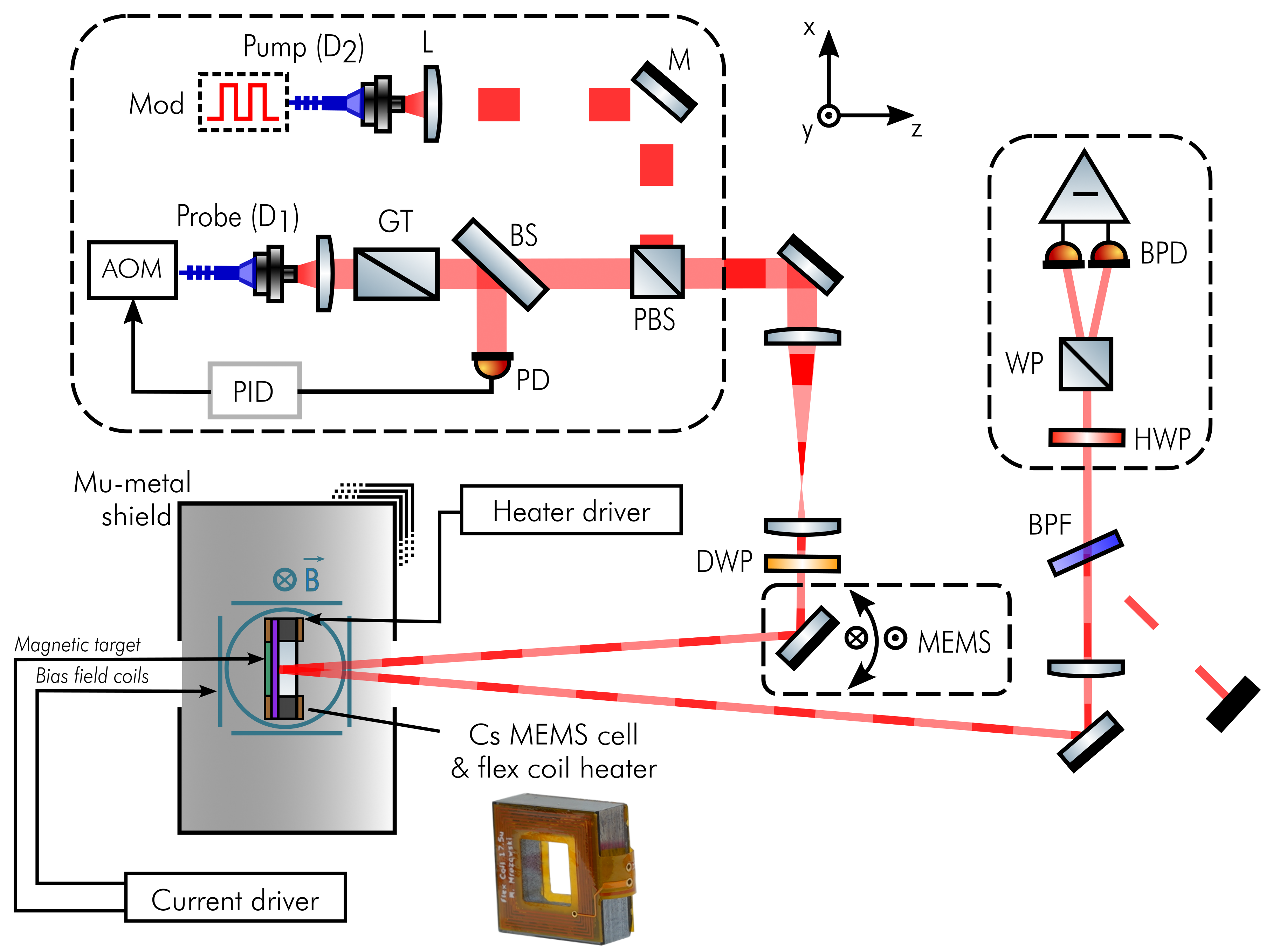}
\caption{Experimental setup for a cesium (Cs) FID magnetometer using co-propagating pump (D$_2$) and probe (D$_1$) beams. The pump is amplitude-modulated, while the probe intensity is stabilized using an acousto-optic modulator (AOM) via feedback to a photodiode (PD) through a Glan–Taylor polarizer (GT) and beam splitter (BS). A polarizing beam splitter (PBS) combines the beams, which are then reduced to a beam waist of around $250\,\mathrm{\mu{m}}$ by a telescope lens (L) system and directed into the Cs micro-electromechanical systems (MEMS) vapor cell, housed within a multi-layer mu-metal shield, via a two-axis MEMS scanning mirror. A dual-wavelength waveplate (DWP) circularly polarizes the pump while maintaining the probe’s linear polarization. After interaction with the cell, the probe passes through a bandpass filter (BPF) and is re-imaged onto a polarimeter consisting of a half-wave plate (HWP), Wollaston prism (WP), and balanced photodetector (BPD). Current and heater drivers independently control the bias magnetic field and cell temperature.}
\label{fig:experimental_setup}
\end{figure}
\indent Figure \ref{fig:experimental_setup} shows a schematic representation of a magnetic field imaging setup based on a FID OPM \cite{grujic2015sensitive, hunter2018free, hunter2018waveform, hunter2023free, hunter2023optical}. The Cs vapor cell contains approximately 220~Torr of nitrogen (N\textsubscript{2}) buffer gas. The silicon cavity dimensions are $(6 \times 6 \times 3)\,$mm; details of the fabrication and pressure calibration methods are described in~\cite{hunter2023free, dyer2023nitrogen}. The vapor cell is $4~$mm thick in total, including the glass interfaces. The buffer gas slows the diffusion of atoms toward the cell walls~\cite{mcwilliam2024optimizing}, thereby reducing wall-collision rates and spatially confining the atomic ensemble; an essential requirement for the imaging technique employed in this work. \\
\indent Assuming no standoff, the spatial resolution would be dependent on the distance the atoms diffuse within the interrogation period. This is known as the diffusion crosstalk-free distance which was estimated to be around $300~\mu$m~\cite{dong2019spin, horsley2015high}. The diameters of the optical interrogation beams were measured to be approximately $250~\mathrm{\mu{m}}$ ($1/e^2$ diameter), set to closely match this diffusion-limited distance~\cite{hunter2023free}. The beam diameters were determined using a CMOS camera positioned at the same distance as the vapor cell via a flip-mirror arrangement. The Rayleigh length ($Z_{R} \approx 5.5~$cm) is over an order of magnitude greater than the cell thickness, thus intensity gradients along the propagation axis are considered to be negligible.  \\ 
\indent In this FID system, the pump beam is pulsed while the probe light is continuous-wave (cw), thereby temporally separating the optical pumping and readout phases of the magnetic measurement. The alkali vapor is optically pumped and probed using separate, co-propagating laser beams at wavelengths of approximately 852.3~nm and 894.6~nm, corresponding to the Cs D$_2$ and D$_1$ lines, respectively. The pump beam is tuned on-resonance with the pressure-shifted $F = 3 \rightarrow F'$ transition and is provided by a grating-stabilized single-frequency diode laser (LD852‐SEV600) capable of delivering up to 600~mW of optical power. The linewidth is small ($\sim 20\,$MHz) compared to the pressure-broadened atomic transition, which has a full width at half maximum of 3.7~GHz due to collisions with the buffer gas. \\
\indent A maximum optical pump power exceeding 70~mW is available at the cell after beam conditioning and a fiber-coupling stage. The pump light amplitude is modulated from its maximum value to effectively zero ($<10~\mu$W) using an acousto-optic modulator. A 1~kHz pulse train is applied with a 10~\% duty cycle. The remaining 90~\% of the cycle is allocated for spin readout performed by the cw probe light. These modulation parameters provide an optimal balance between readout time and spin preparation, minimizing dead time while maintaining sensitivity. This pulsed FID sampling imposes a Nyquist limit of $500\,\mathrm{Hz}$ on magnetic signals measured.
\indent Although both laser systems produce a linearly polarized output, the probe beam passes through a Glan–Thompson polarizer to ensure high polarization purity for detection. The pump and probe beams are combined using a polarizing beam splitter to ensure co-propagation. A telescope system is used to reduce both beam diameters from $2\,\mathrm{mm}$ to $0.25\,\mathrm{mm}$. The probe light is linearly polarized and blue-detuned by approximately 80~GHz from the $F = 4 \rightarrow F' = 3$ transition of the Cs $\mathrm{D_1}$ line, and is generated by a distributed Bragg reflector laser (DBR895PN). This large frequency detuning minimizes power broadening effects, which would otherwise be significant due to the high intensity of the narrow probe beam. A probe power of approximately $1\,\mathrm{mW}$ is used, close to the maximum level at which the sensor remains photon shot-noise limited. To ensure long-term stability, the amplitude of the probe beam is actively stabilized, suppressing light-shift fluctuations caused by intensity drifts. \\
\indent The combined beams are spectrally separated, allowing a dual-wavelength multi-order waveplate to convert the pump beam to circular polarization while preserving the probe beam’s linear polarization. This configuration optimizes the efficiency of both the optical pumping and spin readout processes. Alternative beam geometries, such as using a single elliptically polarized beam~\cite{shah2009spin} or introducing a small angle between the pump and probe beams~\cite{karaulanov2016spin}, have been investigated in prior work; however, these approaches generally reduce magnetometric sensitivity and are therefore not employed here. A spectral bandpass filter is placed before the balanced photodetector to suppress residual pump light and prevent it from contributing to the signal. The resulting signal is digitized using a Picoscope (model 5444D), with data acquired at a downsampled sampling frequency of $1/\Delta{t} = 2$~MHz~\cite{hunter2023free, hunter2023optical}. \\
\indent To improve optical pumping efficiency and facilitate compact integration, the vapor cell is embedded in a custom flexible Kapton PCB as seen in Fig. \ref{fig:experimental_setup}. This flexible PCB serves two purposes. First, a magnetic field ($\sim 4~$mT) parallel to the propagation axis of the optical beam is applied during the pumping phase, synchronized with the optical pulse~\cite{hunter2023optical}. This configuration establishes a well-defined quantization axis during optical pumping such that near-unity spin polarization accumulates in the $F = 4$ hyperfine ground state~\cite{schultze2015improving}. As a result, light-narrowing is exploited to suppress the contribution of spin-exchange collisions to the transverse relaxation rate, $\gamma_2$, enhancing sensitivity~\cite{scholtes2011light}. Secondly, the coils provide thermal control to heat the vapor cell and ensure adequate atomic density. Magnetic noise from the heating circuit is effectively eliminated by switching off the field during the detection phase~\cite{hunter2023optical}.  Compared to the FR-4 PCBs used in~\cite{hunter2023optical}, which have a thickness of 0.8~mm, these coils offer more efficient heating due to improved heat transfer and reduced thermal mass. Notably, this thinner ($\sim 0.1\,$mm) coil heater design lowers the minimum standoff distance from the DUT. \\
\begin{figure}[t]
\centering
\includegraphics[width=\columnwidth]{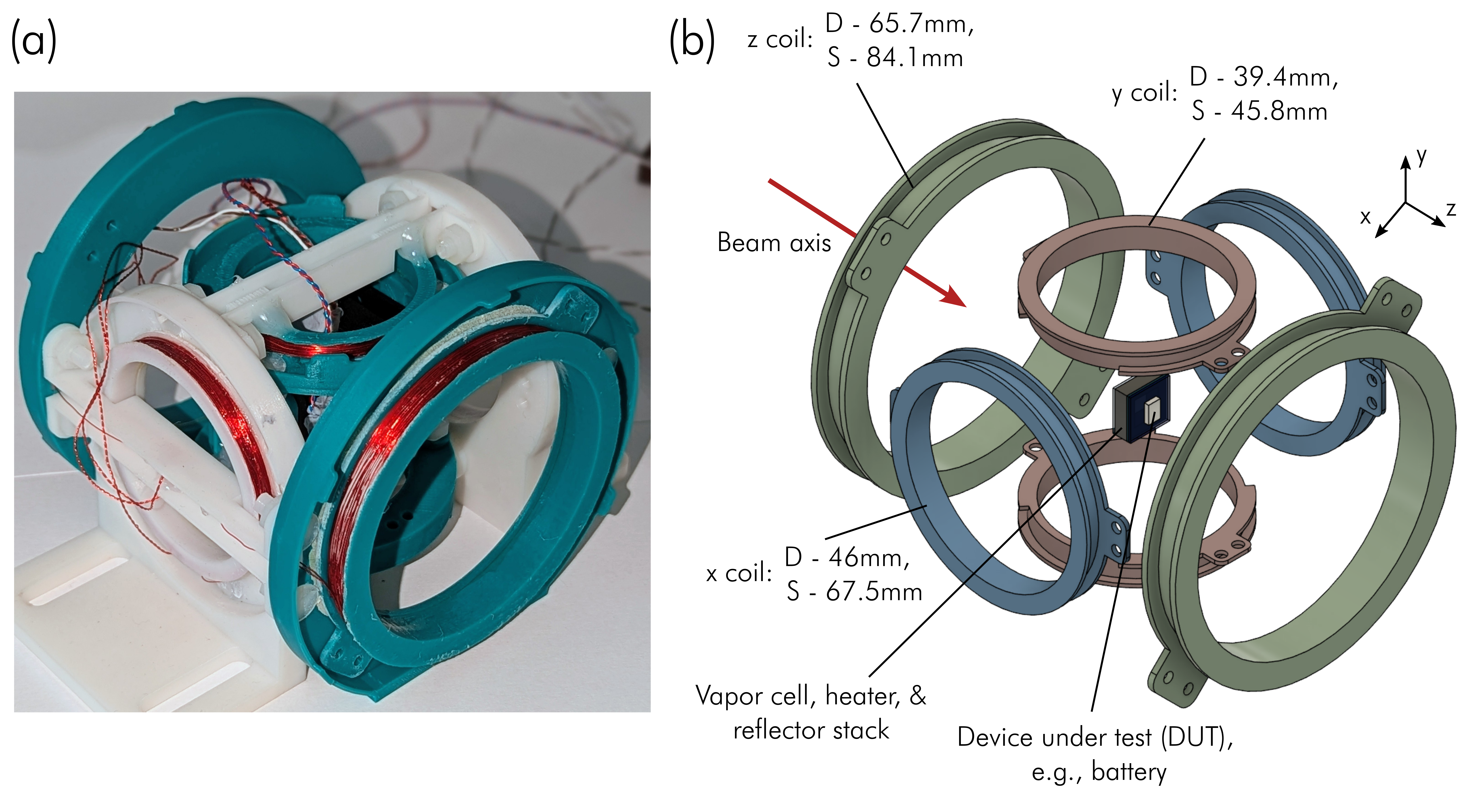}
\caption{(a) Photograph of the experimental assembly inside the mu-metal magnetic shield. (b) CAD model of the assembly illustrating the placement, diameter (D), and separation (S) of the x- (blue), y- (red), and z-axis (green) coil formers. The Cs MEMS vapor cell is mounted at the center of the coil assembly, with the device under test (DUT), such as a surface-mount battery, positioned directly behind the vapor cell.}
\label{fig:coil_assembly}
\end{figure}
\indent The sensor head assembly is enclosed in a four-layer mu-metal magnetic shield (MS-1L, Twinleaf~\cite{Twinleaf_MS1L}), which passively suppresses ambient magnetic interference. The vapor cell is mounted inside a set of custom-made three-axis field coils, as shown in Fig. \ref{fig:coil_assembly}(a), with the relative placement and dimensions of the coil formers illustrated in Fig. \ref{fig:coil_assembly}(b). It can also be seen that the DUTs measured in this work are positioned directly behind the reflector affixed to the vapor cell. The shield also includes integrated magnetic field coils that provide full three-axis magnetic-field control with dedicated windings for first-order gradient compensation. The solenoid coil is aligned along the shield’s axial $x$-axis, while the orthogonal directions use saddle-coil geometries optimized for transverse uniformity. All coils are powered by a highly stable programmable current driver~\cite{mrozowski2023ultra}, whose noise is negligible compared to the intrinsic noise floor of the magnetometer. 

\subsection{Frequency Estimation}
\label{subsection:frequency_estimation}
\begin{figure}[t]
\centering
\includegraphics[width=\columnwidth]{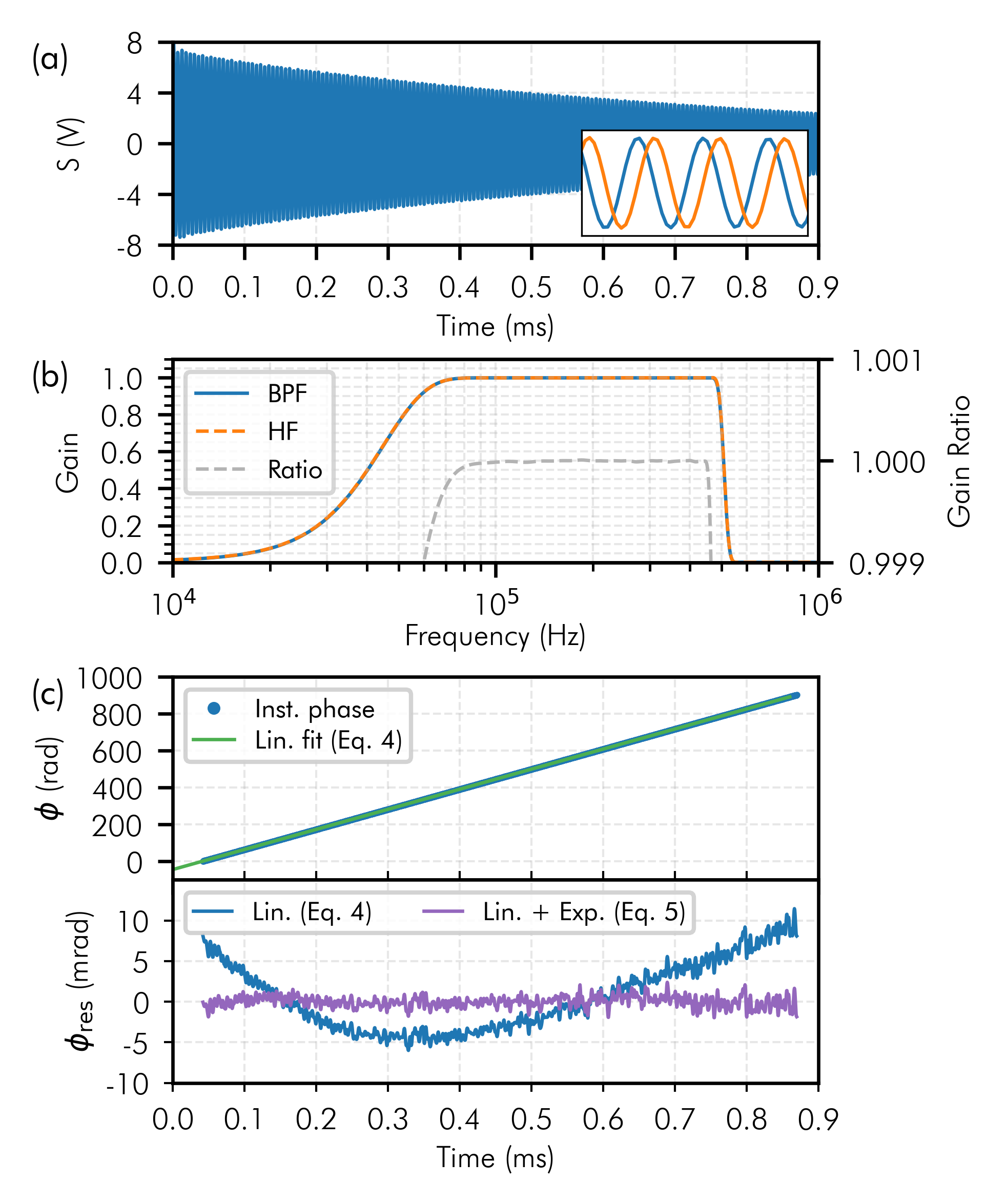}
\caption{Hilbert transform-based frequency extraction technique. (a) Raw FID signal exhibiting damped sinusoidal oscillations at a frequency of approximately $173\,$kHz. Inset: zoomed-in view, denoted by the black dashed lines, showing the in-phase (blue) and quadrature (orange) components extracted after applying finite impulse response (FIR) filters. (b) Frequency response of the bandpass (solid blue) and Hilbert transformer (dashed orange) FIR filters. The gain ratio between the two filters is overlaid in gray on the secondary axis, showing variation within the $\mathrm{70–500\,kHz}$ passband. (c) Top: Instantaneous phase $\phi(t)$ extracted from the argument of the analytic signal (see Eq. \ref{eq:instantaneous_phase}). Bottom: residuals from a linear fit (blue) and from a fit including an additional exponential term (purple).}
\label{fig:hilbert_analysis}
\end{figure}

This subsection outlines the DSP approach used to extract magnetic field information from the FID signals. During the detection phase, the atomic spins undergo Larmor precession at a frequency determined by the magnitude of the local magnetic field 
\begin{equation}
    \omega_{L} = \gamma B_0,
\end{equation}
where $\gamma \approx 3.5$~Hz/nT is the gyromagnetic ratio of Cs. The atomic polarization decays due to several relaxation mechanisms, including collisions with the cell walls, interactions with buffer gas atoms, and spin-exchange collisions among Cs atoms~\cite{mcwilliam2024optimizing, scholtes2014intrinsic}. A typical signal resulting from this precession, taking the form of a damped sinusoid, is depicted in Fig.~\ref{fig:hilbert_analysis}(a). \\
\indent Multiple techniques are available for extracting the Larmor precession frequency, each offering different trade-offs in sensitivity, temporal resolution, and computational efficiency~\cite{zielinski2011frequency}. Fitting the data to a nonlinear damped sinusoidal model has been adopted in previous works~\cite{hunter2018free, hunter2018waveform, hunter2023free, hunter2023optical}, and enables frequency estimation with a precision approaching the Cramér–Rao lower bound (CRLB), which represents the theoretical limit set by the signal-to-noise ratio and damping rate~\cite{gemmel2010ultra}. Here, we demonstrate an alternative method based on a Hilbert transform. \\
\indent The Hilbert transform approach~\cite{wilson2020wide} offers significantly greater computational efficiency, which makes it particularly advantageous for the magnetic imaging strategy employed here by enabling faster scanning rates. Two finite impulse response (FIR) filters are applied independently to the raw FID signals. The first is a Hilbert transformer FIR filter, which imparts a $90\,^{\circ}$ phase shift to all frequency components within a specified passband. The second is a matched bandpass filter with a gain response identical to that of the Hilbert transformer, as illustrated in Fig.~\ref{fig:hilbert_analysis}(b). The outputs of these filters yield the quadrature $Q(t)$ and in-phase $I(t)$ components, respectively, from which the complex analytic signal is constructed as
\begin{equation}
    z(t) = Q(t) + i\,I(t).
    \label{eq:analytic_signal}
\end{equation}
The accuracy of this method depends on the gain ratio, defined as the ratio of the gain responses of the two FIR filters. Within the flat passband, the gain ratio varies by around $0.0005\,\%$, representing negligible systematic uncertainty introduced by filter mismatch. Although the passband can be extended to lower frequencies with comparable gain ratio variation, doing so requires additional filter coefficients, resulting in a longer impulse response. Consequently, more of the initial portion of the FID signal must be truncated, ultimately degrading the achievable sensitivity. \\
\indent The FID signal is linearized by computing the instantaneous phase, defined as
\begin{equation}
    \phi(t) = \arg\{z(t)\} = \mathrm{tan^{-1}}\bigg(\frac{Q(t)}{I(t)}\bigg),
    \label{eq:instantaneous_phase}
\end{equation}
as shown in Fig.~\ref{fig:hilbert_analysis}(c). This can be expressed in terms of the angular Larmor frequency, $\omega_{L}$, and an initial phase offset, $\phi_{0}$, as
\begin{equation}
    \phi(t) = \omega_{L}t + \phi_{0},
    \label{eq:larmor}
\end{equation}
assuming that $\omega_{L}$ is constant. A linear fit applied to the unwrapped instantaneous phase data yields the Larmor frequency from the fitted slope. This approach offers high computational efficiency and can reach the CRLB after accounting for the filter impulse response~\cite{ingleby2022digital}.  \\
\indent The variation in the linear-fit residuals observed in Fig.~\ref{fig:hilbert_analysis}(c) arises from the decaying spin polarization~\cite{hewatt2025investigating}, which perturbs the Larmor frequency during a single FID cycle~\cite{hunter2023optical}. To account for this effect, Eq.~\ref{eq:larmor} can be extended with an exponential correction term giving
\begin{equation}
    \phi(t) = \omega_{0}t + \phi_{0} + A_{p} e^{-\gamma_{p} t},
    \label{eq:larmor_exp}
\end{equation}
where $\omega_{0}$ is the DC component of the time dependent Larmor frequency, $A_p$ is the amplitude of the correction, and $\gamma_p$ the associated decay constant. By evaluating the time derivative at $t = 0$, the corrected Larmor frequency can be calculated as
\begin{equation}
\omega_{L} = \left.\frac{d\phi}{dt}\right|_{t=0} = \omega_{0} - \gamma_{p}A_{p}. 
\end{equation}
In this way, we can obtain the measured Larmor frequency at the instant of highest ground-state polarization~\cite{hewatt2025investigating}. It can be seen from Fig.~\ref{fig:hilbert_analysis}(c) that fitting this extended model considerably reduces residual trends. 

\subsection{Two-Axis MEMS Micromirror Implementation}
\label{subsection:mems_mirror}
\begin{figure}[t]
\centering
\includegraphics[width=0.95\columnwidth]{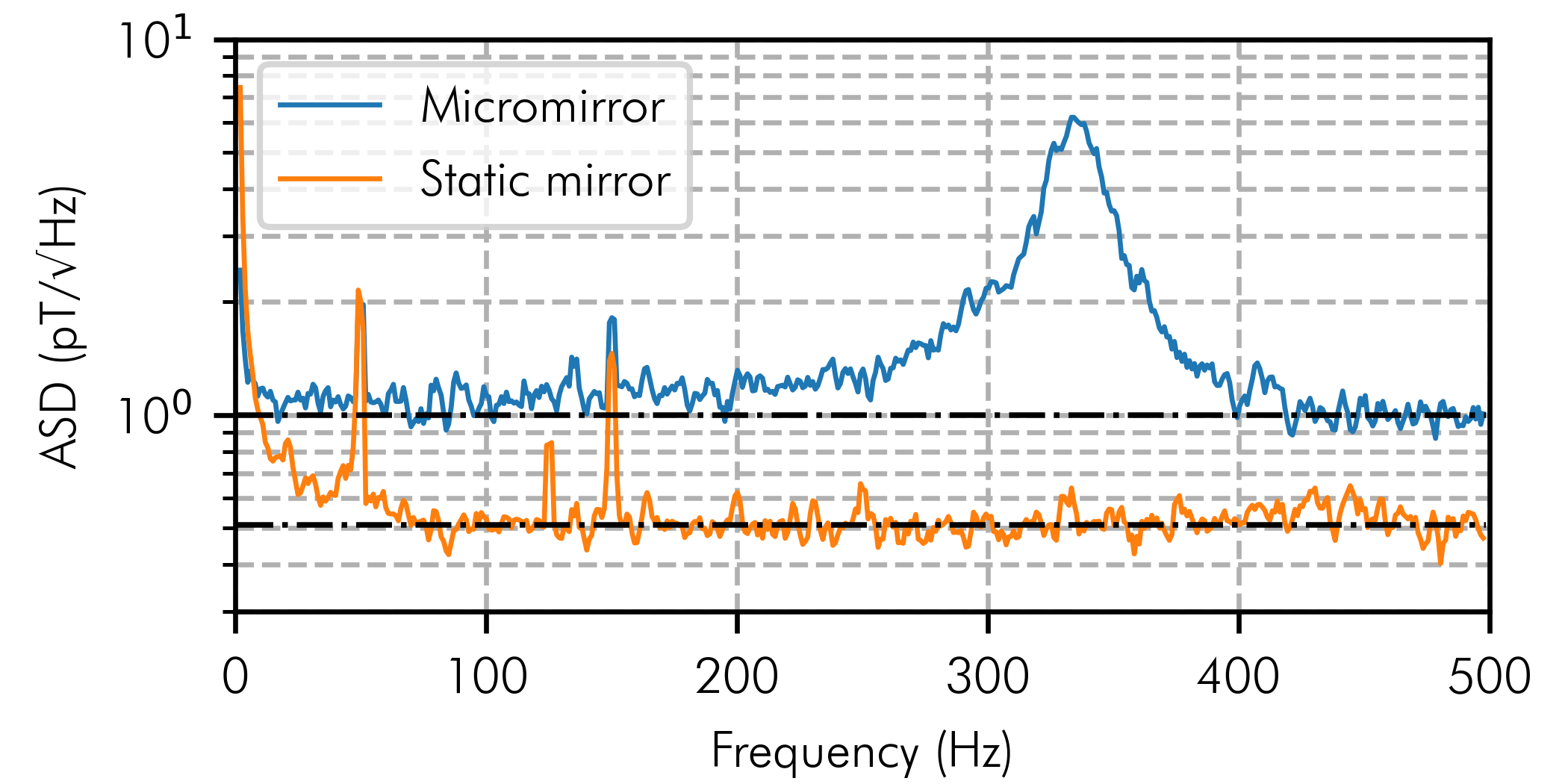}
\caption{Amplitude spectral density (ASD) derived from the time-domain OPM output. The blue and orange traces show a comparison of system performance using a MEMS scanning micromirror versus a standard mirror (Thorlabs BB1-E03), respectively. Both were kept stationary during the measurement. Welch's method was used with 20 non-overlapping segments to reduce spectral fluctuations~\cite{welch1967use}. The static mirror configuration achieves an optimal sensitivity of approximately $0.5\,\mathrm{pT/\sqrt{Hz}}$, while the MEMS mirror exhibits elevated noise due to mechanical resonance near $1.3\,\mathrm{kHz}$, which is folded into the magnetometer bandwidth via aliasing at around $330\,\mathrm{Hz}$. This is caused by coupling between spatial magnetic gradients and mirror motion, and also elevates the broadband noise floor to around $1\,\mathrm{pT/\sqrt{Hz}}$.}
\label{fig:sensitivity}
\end{figure}

Integration of a gimbal-less 2D MEMS micromirror with $2~$mm mirror diameter (Mirrorcle Inc, A7M20.2-2000AL~\cite{mirrorcle2024mems}) for automated beam steering eliminates the need for manual translation, reducing data acquisition times by nearly two orders of magnitude. Magnetic images were acquired using a raster scan approach over an approximately ($4 \times 4$)\,mm field of view and $< 400\,\mu\mathrm{m}$ step size. At each scan position, a train of FID signals over a $100\,\mathrm{ms}$ time period are recorded. \\
\indent The $250\,\mathrm{\mu{m}}$ diameter co-propagating pump and probe beams are directed onto the mirror and subsequently reflected onto the vapor cell. A $0.5\,\mathrm{mm}$ thick reflector is affixed to the rear surface of the cell using an optical adhesive, enabling a double-pass geometry. Compared to the single-pass configuration used in Ref.~\cite{hunter2023free}, this approach offers two key advantages: (i) the optical rotation signal is effectively doubled due to the increased interaction length, and (ii) the DUT can be positioned directly behind the reflector without obstructing the beam path, thereby minimizing and better defining the standoff distance. To facilitate detection, the probe and pump beams were aligned at a slight angle, directing the transmitted light onto the photodetector. This alignment reduces the effective interrogation area to approximately $(4\times4)$~mm under optimal conditions. After transmission through the vapor cell, the light is directed onto a lens that re-images the beam onto a balanced photodetector, which helps prevent beam clipping during scanning. \\
\indent The micromirror exhibits a mechanical resonance at approximately $1.3\,\mathrm{kHz}$ in both movement axes~\cite{mirrorcle2024mems}. As a result, its step-response performance is governed by an external low-pass filter in the driving electronics~\cite{mirrorcle2022easydrive}, typically set to half the resonant frequency to suppress mechanical ringing. Although the magnetometer bandwidth is Nyquist limited to $500\,\mathrm{Hz}$, residual mechanical ringing from the micromirror introduces aliased noise centered near $330\,\mathrm{Hz}$. \\
\indent Figure~\ref{fig:sensitivity} compares the magnetometer’s sensitivity when using the micromirror versus a standard static mirror (Thorlabs, BB1-E03). The prominent noise peak at $330\,\mathrm{Hz}$ results from coupling between spatial magnetic field gradients, e.g., due to bias field inhomogeneity, and the mirror’s mechanical resonance, leading to spurious modulation of the magnetometer signal. This was confirmed using two separate methods. The first included increasing the strength of the magnetic field gradient, which in turn elevated both the noise peak amplitude and the overall noise floor. The broadband noise level is increased due to the spectral wings of the mechanical resonance, limiting the sensitivity to approximately $1\,\mathrm{pT}/\sqrt{\mathrm{Hz}}$ when optimally configured. In the second case, the micromirror was replaced with a standard stationary mirror, while all other experimental parameters were kept constant. Under these conditions, a noise floor of $0.5\,\mathrm{pT}/\sqrt{\mathrm{Hz}}$ was achieved, and the noise peak previously centered at $330$~Hz was no longer observed as seen in Fig.~\ref{fig:sensitivity}. The CRLB predictions~\cite{hunter2023optical}, based on the FID signal characteristics for both mirror configurations, yield a sensitivity limit of approximately $0.5\,\mathrm{pT}/\sqrt{\mathrm{Hz}}$. This result further confirms the impact of vibrationally induced noise from the micromirror on the broadband sensor sensitivity. \\
\indent To further mitigate micromirror-induced ringing~\cite{mirrorcle2024mems}, the low-pass filter cutoff frequency in the micromirror driving electronics was reduced from the default value of $500\,\mathrm{Hz}$ to $100\,\mathrm{Hz}$. This was achieved by adjusting the system clock frequency of the driving electronics, which directly sets the effective filter cutoff frequency~\cite{mirrorcle2022easydrive}. Whilst filtering reduces the mirror-induced noise, complete suppression was not achievable. Nonetheless, this did not pose a limitation for the measurements presented here, as the DUTs investigated exhibited dynamics well below the imposed bandwidth limit. \\
\indent The magnetic field distributions presented in this work were acquired using a raster scanning procedure to systematically map a grid of the spatially varying magnetic field. At each grid point, the Larmor frequencies were extracted from a train of FID signals recorded over a user-defined acquisition period, typically set to $100\,\mathrm{ms}$ to balance per-point precision with overall scan-time efficiency, and averaged to yield a single magnetic field value~\cite{hunter2023free}. The precision at each grid point is governed by the acquisition time and can be adjusted to meet the contrast requirements necessary to resolve the magnetic features of interest. Increasing the acquisition time per point improves precision; however, also extends the total scan duration. \\
\indent Another factor influencing the overall imaging speed is the signal processing time required to extract the Larmor frequency from each FID trace. When using a nonlinear least-squares fitting approach~\cite{hunter2018free}, the computation time varies depending on convergence conditions and constitutes the dominant contribution to the total scan time. In contrast, the Hilbert transform method provides over an order-of-magnitude improvement in processing efficiency and offers more consistent timing performance. \\
\indent Consequently, the minimum time required to complete a full scan is determined by three factors: the micromirror step response time, the total number of grid points used to construct the image, and the magnetometer acquisition and processing time, with the latter being the dominant contribution in the present work. Importantly, the total number of grid points should also be selected with consideration of the system’s spatial resolution limits, which are constrained by atomic spin diffusion within the vapor cell, beam waist, and the standoff distance between the DUT and the sensing volume \cite{hunter2023free}. These physical factors define the smallest magnetic features that can be resolved and thus guide the appropriate grid density. By optimizing the mirror control, signal processing, and spatial sampling parameters, scan durations can be minimized while maintaining high-fidelity magnetic imaging.

\section{Results}
\subsection{Magnetic Gradient Analysis}
\begin{figure}
\centering
\includegraphics[width=0.95\columnwidth]{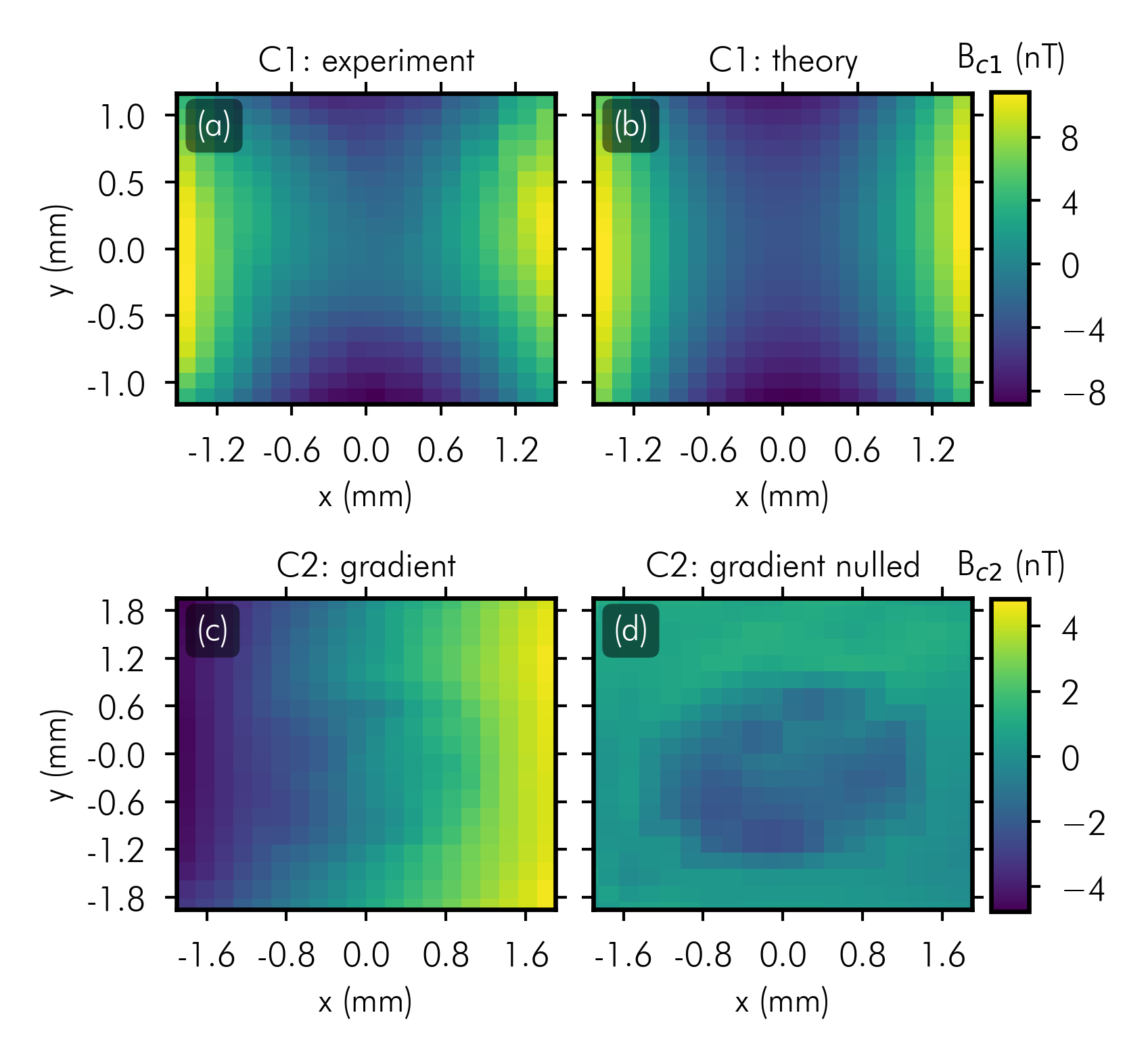}
\caption{Magnetic field inhomogeneities produced by two separate bias-field coil configurations. (a) Measured and (b) simulated magnetic field profiles for a custom-made circular coil pair (C1), separated by $67.5$~mm and having a diameter of $46$~mm, each comprising 65 turns. The drive current was $5.1$\,mA ($\approx 3.2~\mu$T). (c) Measured and (d) gradient-nulled magnetic field distributions using the integrated solenoidal coil in the MS-1L magnetic shield (C2), with specifications provided in \cite{Twinleaf_MS1L}, at a drive current of $50.1$\,mA ($\approx 6~\mu$T). The field shown in (d) results from applying the integrated gradient cancellation coils to suppress the dominant first-order components observed in (c).}
\label{fig:magnetic_gradients}
\end{figure}

The performance of the imaging system was first evaluated through characterization of the applied bias field and magnetic-field uniformity within the shielded region. This provides a quantitative benchmark for validating the field reconstruction accuracy of the system before examining more complex current-carrying devices. \\
\indent Figure~\ref{fig:magnetic_gradients} presents an analysis of the magnetic-field distributions generated by two distinct bias-field coil configurations within the magnetic shield, with the bias field applied along the $x$-axis in each case. The custom-made coil pair (C1) is constructed from two 46-mm-diameter formers separated by 67.5 mm, as shown in Fig. \ref{fig:coil_assembly}(b), with 65 turns per coil and a drive current of 5.1 mA. The coil parameters were selected to achieve a high field gain, thereby minimizing the required drive current, while the resulting field nonuniformity provided a useful test case for validating the imaging performance. While the FID magnetometer performance is largely insensitive to the absolute bias-field magnitude~\cite{hunter2023optical}, this parameter influences the degree of field non-uniformity within the measurement volume. The chosen bias level was therefore selected to provide a modest spatial gradient across the imaging region. \\
\indent Based on these coil parameters, the magnetic image obtained using the proposed imaging system is compared with the expected field profile, shown in Figs.~\ref{fig:magnetic_gradients}(a) and (b), respectively. The simulated distribution was obtained using the Magpylib software package for magnetic-field computation \cite{Tiercelin2022magpylib}, which evaluates the Biot–Savart law at the optical-beam position over a two-dimensional grid to generate the theoretical reference field map shown in Fig.~\ref{fig:magnetic_gradients}(b). To emphasize spatial gradients, the mean magnetic field over each measurement grid was subtracted, following the procedure described in \cite{hunter2023free}. The measured field distribution closely matches the simulated pattern, confirming both the calibration accuracy of the imaging system and the validity of the coil-field model. \\
\indent A small angular dependence is observed in the measured field map, which is attributed to mechanical tolerances in the mounting of the vapor cell within the coil former. Introducing a $1.5^{\circ}$ tilt into the simulated geometry reproduces this behavior, resulting in close agreement between the measured and calculated magnetic-field distributions. This observation indicates that the residual discrepancies primarily arise from minor mechanical misalignment rather than modeling error. \\
\indent The magnetic field produced by the custom-made coil pair (C1) exhibits higher-order components beyond the first spatial derivative, which complicates the generation of a uniform bias field through simple gradient cancellation. Therefore, a second configuration (C2) employing the integrated coil system of the MS-1L magnetic shield was evaluated~\cite{Twinleaf_MS1L}. The coil gains for configurations C1 and C2 are $634\,\mathrm{nT/mA}$ and $130\,\mathrm{nT/mA}$, respectively. However, the C2 configuration generates a substantially more uniform field that is easier to condition. \\
\indent Figure~\ref{fig:magnetic_gradients}(c) shows the measured magnetic field obtained using the solenoid coil, and Fig.~\ref{fig:magnetic_gradients}(d) presents the corresponding distribution after first-order gradient cancellation. The integrated coil system effectively nulls the axial gradient, yielding a highly uniform bias field within the measurement region. The small residual background arises primarily from condensed Cs on the vapor-cell windows, which partially obstructs the optical path and introduces two systematic effects: (1) modified optical-pumping dynamics and (2) localized variations in the fictitious field generated by light shifts. These systematic offsets can be compensated through interleaved background measurements, as described in~\cite{hunter2023free}. These results establish a calibrated and spatially uniform bias-field environment for subsequent magnetic-imaging measurements.

\subsection{Custom Current Configuration}
\begin{figure}
\centering
\includegraphics[width=\columnwidth]{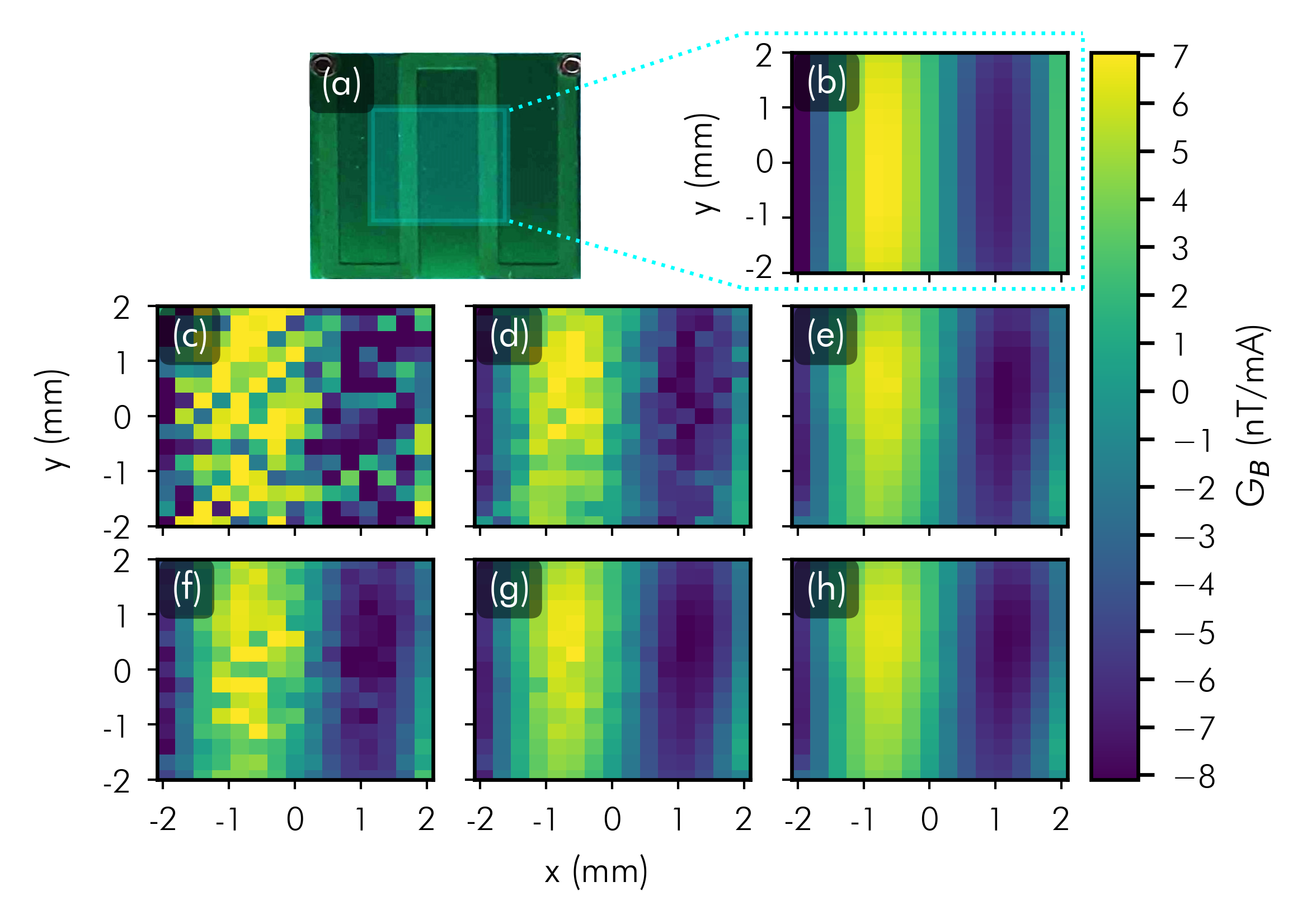}
\caption{(a) Photograph of a custom-made PCB containing alternating-direction copper tracks used as a magnetic field source. The PCB surface dimensions are ($7.5 \times 7.5$)\,mm, with the imaged region highlighted in cyan. The vertical copper tracks are spaced $2\,$mm apart. (b) Predicted magnetic field distribution from this current configuration using the Biot–Savart law~\cite{Tiercelin2022magpylib}. (c–e) Measured ($15 \times 15$) magnetic field grid using constant currents of $\mathrm{2\,\mu{A}}$, $\mathrm{10\,\mu{A}}$, and $\mathrm{100\,\mu{A}}$, from left to right, respectively. The sensor background has been subtracted. (f–h) Corresponding field maps obtained using an alternating current of the same root-mean-squared (rms) amplitudes and a frequency of $111\,$Hz. Magnetic time series captured at each grid point are processed using demodulation as in Ref.~\cite{hunter2023free}, thus no background subtraction is required. All data has been scaled by the applied current, yielding the magnetic gain parameter $G_{B}$ in units of $\mathrm{nT/mA}$.}
\label{fig:copper_track_image}
\end{figure}

A custom current-carrying target was measured to demonstrate imaging performance under well-defined magnetic conditions for a known source geometry. In this case, the DUT was a PCB containing alternating-direction copper tracks, shown in Fig.~\ref{fig:copper_track_image}(a). The PCB has surface dimensions of ($7.5 \times 7.5$)\,mm, and the vertical copper tracks are spaced $2\,$mm apart. The effective standoff distance of the DUTs measured in this work is $2.7\,$mm, determined by the combined thickness of the optical and structural layers separating the DUT from the center of the vapor cell, heater, and reflector stack. The positioning of the DUT with respect to the vapor cell and coil assembly is illustrated in Fig.~\ref{fig:coil_assembly}(b). \\
\indent The predicted magnetic-field distribution, computed using the Biot–Savart law~\cite{Tiercelin2022magpylib} and shown in Fig.~\ref{fig:copper_track_image}(b), exhibits alternating bright and dark regions corresponding to areas where the field from the copper tracks is aligned or anti-aligned with the static bias field along the $x$-axis. All field maps in Figs.~\ref{fig:copper_track_image}(b)–(h) are normalized by the applied current, defining the magnetic-gain parameter $G_B$ in units of $\mathrm{nT/mA}$. \\
\indent The central row of Fig.~\ref{fig:copper_track_image} presents images acquired under direct-current (DC) excitation, and the bottom row shows those obtained using alternating-current (AC) excitation at $111$ Hz, chosen to shift the signal above dominant $1/f$ technical noise. Each column corresponds to current amplitudes of $\mathrm{2~\mu A}$, $\mathrm{10~\mu A}$, and $\mathrm{100~\mu A}$, from left to right, selected to span the magnetometer’s sensitivity range from near the noise floor to high-contrast operation. At $\mathrm{100~\mu A}$, the spatial field pattern is clearly resolved under both DC and AC conditions and agrees closely with theoretical predictions. The slight asymmetry along the $y$-axis is attributed to wiring connections on the PCB. As expected, image contrast decreases with lower current amplitudes; however, the AC measurements retain stronger contrast at $\mathrm{2~\mu A}$ because of the suppression of low-frequency technical noise (see Fig.~\ref{fig:sensitivity}). At higher modulation frequencies, the signal lies above the dominant noise band, and the image contrast is primarily limited by the intrinsic noise floor of the sensor. \\
\indent A key performance metric of the imaging system is its field sensitivity, which is primarily limited by photon shot noise. The achievable spatial resolution is fundamentally constrained by the beam size and by spin-diffusion dynamics within the vapor cell. In practice, magnetic-field roll-off across the optical path imposes an additional technical limitation that could be mitigated by employing thinner vapor-cell geometries. The current optical interrogation area of approximately $(4\times4)$~mm could also be expanded to increase the effective field of view and leverage the pixel density afforded by the MEMS micromirror device. 

\subsection{Functional DUTs: bridge rectifier and ceramic battery}
\begin{figure}[t]
\centering
\includegraphics[width=0.95\columnwidth]{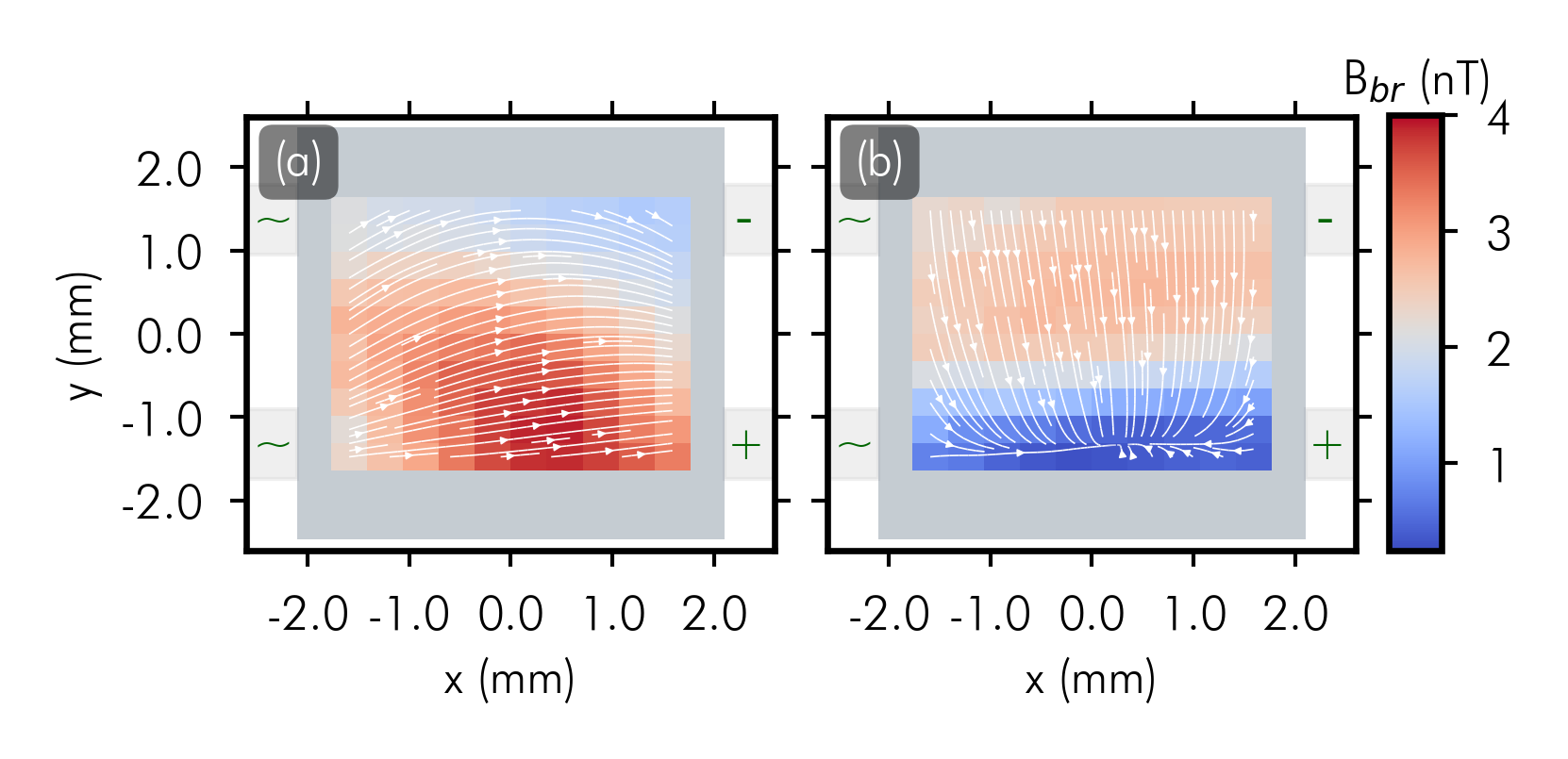}
\caption{Magnetic-field streamlines measured from a bridge rectifier integrated circuit (IC) under opposite current polarities. (a) Measured vector field overlaid on the magnetic-field-magnitude map $B_{br}$ for negative current polarity. (b) Corresponding measurement for positive current polarity. The color scale indicates the local magnetic-field amplitude, while white streamlines depict the direction of the in-plane magnetic field. As the OPM is a scalar sensor, the $x$- and $y$- components were determined by applying the bias field along the corresponding axis and performing independent measurements. The asymmetry between the two cases suggests that different internal current pathways are activated under forward and reverse bias, consistent with diode-based rectification behavior. The IC is represented by the grey outline showing the two AC input terminals ($\sim$) and the positive and negative DC output terminals. The surface dimensions of the IC are ($4.2 \times 4.95$)~mm.}
\label{fig:bridge_rectifier_image}
\end{figure}

\begin{figure}[t]
\centering
\includegraphics[width=\columnwidth]{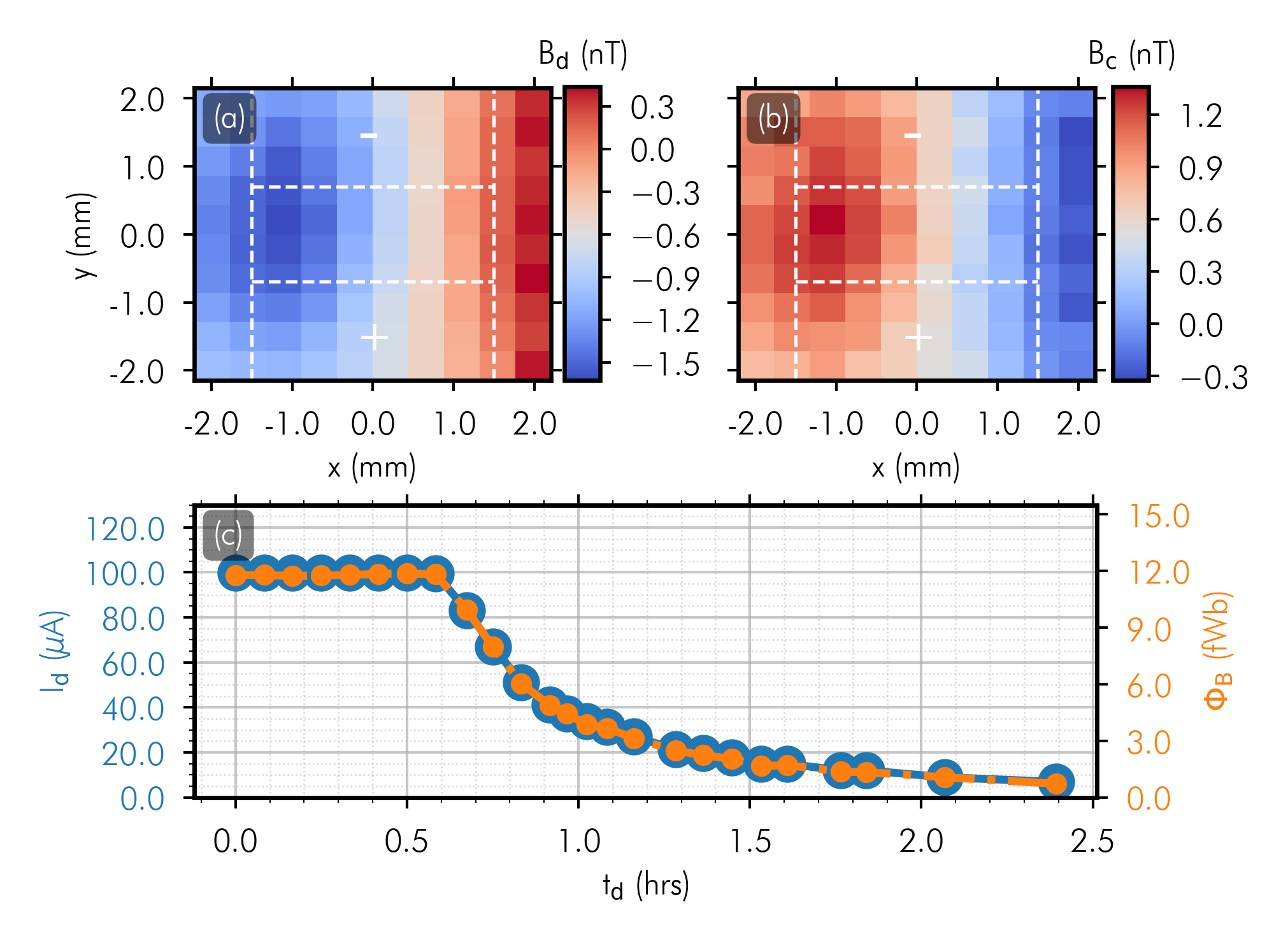}
\caption{Magnetic-field images showing the $x$-axis spatial component for current flow in a ceramic battery under different operating conditions. The battery is connected to external circuitry via a twisted copper-wire pair oriented along the $z$-axis. (a) Magnetic-field distribution $B_d$ during discharging with a steady current of $100~\mu\mathrm{A}$ until the battery is unable to sustain this current. (b) Magnetic-field distribution $B_c$ during charging using an external power source. The color maps show the net field after subtraction of the sensor background, as in previous instances. (c) Measured discharge current $I_d$ (blue, left axis) and integrated magnetic flux $\Phi_B$ (orange, right axis) as a function of discharge time $t_d$, demonstrating the time evolution of battery output as it depletes. The battery outline is depicted by white dashed lines, with polarity indicated as well. The surface dimensions of the battery are ($4.4 \times 3$)~mm.}
\label{fig:battery_image}
\end{figure}

\indent Having validated the imaging system using a well-defined current geometry, the subsequent experiments were conducted on functional electronic devices to demonstrate its applicability to practical IC and surface-mount battery components. These DUT measurements were acquired using the same raster-scanning and frequency-extraction protocol, described in Sections~\ref{subsection:frequency_estimation} and~\ref{subsection:mems_mirror}. \\
\indent Figure~\ref{fig:bridge_rectifier_image} shows the reconstructed magnetic-field distributions produced by a bridge rectifier IC (Onsemi, MB8S) under opposite current-biasing conditions. Spatial maps of the in-plane field magnitude $B_{br}$ are overlaid with white streamlines indicating local field direction. These maps were reconstructed by combining scalar OPM measurements obtained with the bias field aligned along the $x$- and $y$-axes, respectively, enabling visualization of both $B_x$ and $B_y$ components. \\
\indent In a bridge rectifier, the current polarity determines which internal diode pairs conduct, thereby altering the internal current pathways and the resulting magnetic-field distribution. For negative current polarity [Fig.~\ref{fig:bridge_rectifier_image}(a)], the measured field exhibits a dipole-like structure with pronounced curvature of the magnetic-field streamlines, indicative of spatially confined current flow through an asymmetric internal layout. Reversing the current polarity [Fig.~\ref{fig:bridge_rectifier_image}(b)] changes the overall field direction; however, the distribution is not simply inverted. Instead, differences in both field magnitude and streamline curvature are observed, with regions of enhanced curvature under negative polarity, suggesting a more confined current path compared to the broader distribution under positive bias. This polarity-dependent asymmetry is consistent with the activation of distinct internal diode pairs and non-symmetric routing commonly employed in compact rectifier packages. Although the internal metallization is not directly accessible, the measured magnetic-field maps provide indirect evidence of current redistribution within the device. \\
\indent Having confirmed the capability to resolve polarity-dependent current pathways within an IC, the technique was next applied to an energy-storage system to evaluate dynamic charging and discharging conditions. The DUT was a multilayer ceramic battery (TDK Electronics, BCT1812M101AG) with a nominal voltage of $1.5$~V and a capacity of $100~\mu$Ah. Although its energy density is lower than that of rare-earth batteries of comparable size, the all-solid-state architecture provides reliable, surface-mount operation suited to low-power applications such as internet-of-things (IoT) and wearable devices~\cite{bason2022non}. \\
\indent Figures~\ref{fig:battery_image}(a) and (b) show magnetic-field images recorded during discharging and charging, respectively, with the bias field applied along the $x$-axis. The negative and positive terminals of the battery were oriented to the top and bottom, respectively. The battery was connected using twisted wires to minimize magnetic interference, and background subtraction was performed using interleaved measurements of an electrically neutral state. \\
\indent During discharging, a steady current of $100~\mu\mathrm{A}$ generated the magnetic-field distribution $B_d$ shown in Fig.~\ref{fig:battery_image}(a), while charging under constant-voltage operation produced the inverted pattern $B_c$ in Fig.~\ref{fig:battery_image}(b). The $100~\mu\mathrm{A}$ current amplitude was chosen to enable direct comparison with the reference PCB measurements [Fig.~\ref{fig:copper_track_image}(e)]. The polarity of the measured field is consistent with current reversal between the two operating states. Figure~\ref{fig:battery_image}(c) shows the discharge current $I_d$ together with the corresponding integrated magnetic flux $\Phi_B$ as a function of time. The close correspondence between $\Phi_B$ and $I_d$ throughout the discharge cycle demonstrates the quantitative fidelity of the imaging system for dynamic current monitoring, indicating that the measured magnetic signal is dominated by the primary current path within the battery with minimal contribution from parasitic wiring fields. For a discharge current of $100~\mu\mathrm{A}$, the measured magnetic flux agrees closely with the value of $13.8$~fWb obtained from the reference PCB configuration [Fig.~\ref{fig:copper_track_image}(e)], validating the magnetometer’s quantitative response. \\
\indent The combination of optical beam steering, fast Hilbert-transform-based signal processing, and gradient-compensated bias-field control enables accurate reconstruction of in-plane magnetic-field distributions with sub-picotesla sensitivity and millimeter-scale spatial resolution under the present experimental conditions. In the current implementation, the spatial resolution is primarily limited by the effective standoff distance between the magnetic source and the sensing volume, as well as by the spatial decay of the source field, which is inherently DUT-dependent. \\
\indent Experimental validation of millimeter-scale resolution was demonstrated by imaging a custom PCB containing vertical copper tracks spaced $2$~mm apart, which are clearly resolved in the measured magnetic-field maps. Based on the measured optical beam waist ($\sim250~\mu$m) and the estimated diffusion-limited length scale ($\sim300~\mu$m), the imaging architecture is intrinsically capable of sub-millimeter spatial resolution. Further reduction of the standoff distance through thinner vapor-cell and reflector geometries would enable operation closer to these fundamental limits. Additionally, increasing buffer-gas pressure reduces the spin-diffusion length and improves coherence times through suppression of wall-collision effects. 

\section{Conclusion}
We have demonstrated a fully automated magnetic imaging system based on a FID OPM integrated with a two-axis MEMS scanning micromirror. The system employs a double-pass optical geometry that minimizes the standoff distance between the vapor cell and the DUT to $2.7\,$mm, thereby enhancing spatial resolution by capturing stronger near-field signals before significant magnetic-field decay. \\
\indent The performance of the imaging system was evaluated through characterization of the magnetic-field uniformity of an applied bias field, establishing a quantitative benchmark for validating field reconstruction accuracy. Subsequent measurements analyzed the sensor contrast using a custom PCB with alternating current paths, followed by implementation on a bridge-rectifier to resolve polarity-dependent asymmetries from current pathways within the IC, and on a surface-mount ceramic battery to capture dynamic current evolution during charge and discharge conditions. These measurements validate the system’s suitability for noninvasive diagnostics and quality assurance of ICs and batteries. \\
\indent A computationally efficient Hilbert-transform-based signal-processing approach enables rapid frequency estimation across spatial grids, offering substantial speed improvements over traditional nonlinear fitting. An optimal magnetometer sensitivity of $0.5~\mathrm{pT}/\sqrt{\mathrm{Hz}}$ was achieved using a standard static mirror, with the dynamic micromirror configuration maintaining broadband performance at a slightly elevated level of $1~\mathrm{pT}/\sqrt{\mathrm{Hz}}$ which could be suppressed with improved driving electronics. The technique produces high-precision magnetic images within minutes; representing an improvement of roughly two orders of magnitude in acquisition speed relative to manual translation methods~\cite{hunter2023free}. \\
\indent The achieved sensitivity and spatial resolution are well suited to applications involving current densities on the order of $1~\mathrm{\mu A/mm}$ in low-current sources where other methods either lack sufficient field sensitivity (e.g., NV magnetometers) or are restricted by technical constraints such as the requirement for zero-field operation (e.g., SERF OPMs), which may be difficult to achieve in industrial environments. \\
\indent Compared with alternative magnetic-imaging techniques such as superconducting quantum interference device (SQUID) microscopy, NV-diamond magnetometry, or RF-based OPMs, the present approach offers several advantages. It operates above room temperature without cryogenic cooling, employs nonmagnetic optical beam steering instead of mechanical translation, and directly provides quantitative field amplitudes rather than relative contrast. Additionally, this device provides excellent precision and accuracy in finite-field environments which facilitates unshielded measurement capability.  \\
\indent Despite these strengths, there are potential routes to improvement. The current system relies on sequential field-axis measurements to reconstruct vector maps, which increases acquisition times. Future implementations could employ vectorization protocols to obtain these components simultaneously~\cite{bulatowicz2023feedback}. Integration of real-time frequency readout using field-programmable gate arrays~\cite{gong2025high} could further reduce acquisition latency and enhance the overall bandwidth. Overall, the results establish a practical route toward compact, high-sensitivity and spatially resolved magnetic imaging based on alkali-vapor magnetometry. \\
\indent Future work will aim to further reduce acquisition time, enhance spatial resolution, and extend operation to more dynamic environments. One promising direction involves replacing the MEMS mirror with a digital micromirror device (DMD)~\cite{torre2025microscopic}, which offers faster switching and the potential to explore compressed-sensing strategies for accelerated image acquisition. \\
\indent The data presented in this paper are available from the University of Strathclyde data archive~\cite{hunter2025High}.

\section*{Acknowledgments}

D. Hunter gratefully acknowledges funding support from a Royal Academy of Engineering UK Intelligence Community Fellowship. J. P. McGilligan gratefully acknowledges funding from a Royal Academy of Engineering Research Fellowship. This work was supported by the Engineering and Physical Sciences Research Council under grants EP/W026929/1, EP/Z533166/1, and EP/Z533178/1. T. S. Read gratefully acknowledges support by the National Institute of Biomedical Imaging and Bioengineering of the National Institutes of Health under Award Number U01EB028656. Sandia National Laboratories is a multimission laboratory managed and operated by National Technology \& Engineering Solutions of Sandia, LLC, a wholly owned subsidiary of Honeywell International Inc., for the US Department of Energy National Nuclear Security Administration under contract DENA0003525. The content of this publication is solely the responsibility of the authors and does not necessarily represent the official views of the National Institutes of Health, Sandia National Laboratories, or US Department of Energy. The authors acknowledge funding support from INMAQS.

\ifCLASSOPTIONcaptionsoff
  \newpage
\fi

\bibliographystyle{IEEEtran}
\bibliography{references}











\end{document}
